\documentclass[preprint,aps,floats,nofootinbib,showpacs]{revtex4}
\usepackage{epsfig,epsf}
\usepackage{amssymb}
\usepackage{amsmath}
\usepackage{graphicx}
\newcommand{\be}{\begin{equation}}
\newcommand{\ee}{\end{equation}}

\newcommand{\roughly}[1]%
{\mathrel{\raise.4ex\hbox{$#1$\kern-.75em\lower1ex\hbox{$\sim$}}}}

\newcommand\beq{\begin{eqnarray}}
\newcommand\eeq{\end{eqnarray}}

\def\Dsl{\,\raise.15ex \hbox{/}\mkern-12.8mu D}

\newcommand\MeV{{\rm MeV\,}}

\def\fm3{fm$^{-3}$}

\begin{document}
%
\preprint{\vbox{\hbox{TPJU-7-2006}}}

\bigskip
\bigskip

\title{Comparison of the non-uniform chiral and 2SC phases at finite temperatures and densities}

\author{Mariusz Sadzikowski}

\affiliation{Smoluchowski Institute of Physics,
Jagellonian University, Reymonta 4, 30-059 Krak\'ow, Poland}

\begin{abstract}
We study the phase diagram of the strongly interacting matter at finite temperatures
and densities including 2SC, uniform chiral and non-uniform chiral phases within
the Nambu - Jona-Lasinio model in the mean field approximation.
\end{abstract}
\maketitle

\section{Introduction}

During the last decade a great effort has been made to
understand the structure of matter at finite temperature and density
\cite{reviews}. As a generally accepted hypothesis the strongly
interacting matter at high baryon density is in a color superconducting state. Although the
first attempts in this direction had been made already long ago
\cite{bailin} only recently the subject achieved a new boost
\cite{alfshur}.  The simple observation was made that the
quark-quark interaction at moderate densities should be of the same
order as quark-antiquark interaction. This leads in turn to the
sizable value of the superconducting gap $\Delta\sim O(100)\,\MeV $. What is more
interesting it was proved that because of the non-abelian structure
of the underlying fields in QCD, the superconduction gap grows with the chemical potential
 at very high density \cite{son}.

However, the main interest is
concentrated at moderate densities possible to exist in the core of the compact stars.
Using the selfconsistent calculation in the three flavor Nambu - Jona-Lasinio model
the phase diagram of matter was elaborated under the reasonable assumptions \cite{bublashke}.
It occurs that depending on the value of the diquark coupling constant the 2SC
(or g2SC) phase occupies the large region of the phase diagram at moderate densities and temperatures.
In the construction of the phase diagram the superconducting phases
are usually compared to the uniform chiral phase. However, there are also other possibilities.

It was shown that the chiral phase can develop the wave in
the Lorentz scalar - pseudoscalar channels which has a lower energy than the uniform state \cite{dautbron}.
This effect is well known in the low dimensional models (eg. \cite{thies}). Nevertheless, it can be argued that
the phenomenon of the chiral wave might also appear in the 3+1 dimensional QCD.
The Nambu realization of the chiral symmetry breaking \cite{nambu} is based on the idea borrowed
from the BCS theory of superconductivity. The effective contact four-fermion
interaction between quarks, described by the coupling constant $G$, is assumed to be attractive
in the scalar, isoscalar channel ($G>0$). It is then energetically favorable for the system to
create the quark-antiquark pairs which built the Lorentz scalar chiral condensate.
However QCD is invariant under $SU(2)_V\times SU(2)_A\times U(1)$ symmetry. As a result
the effective quark interaction also contains pseudoscalar, isovector channel with the same
coupling constant $G$. If the chiral wave develops in both channels then, as one can show,
the wave vector acts effectively as a constant magnetic field which couples to the spin of
the quarks. Then, in the analogy to the Zeeman effect, the quark energy levels split.
The half of the energy branches lower their energy in compare to the uniform case,
which depends on the quark spin projection on the wave vector axis. At finite density the
quarks built the Fermi ladder, first filling the branches of the lower energy. This energy
gain competes with the energy price due to the field gradients. There is a question of
the detail calculations which contribution is more important. It was shown within the
framework of the Nambu - Jona-Lasinio models or linear sigma models \cite{dautbron, sad1} that there is a
wave vector range in which the energy gain overcome the price. The above scenario shares
similarity with the Overhauser effect \cite{overhauser} which was already pointed out in
the paper \cite{nakano}.



As one can see it is an important task to compare 2SC phases with non-uniform
rather than uniform chiral phase.
As a first attempt the 2SC phase was compared to the non-uniform chiral phase at zero
temperature \cite{sad2}. In this paper the calculation was extended to non-zero temperature.
As a result the phase diagram in $T-\mu$ plane of the two flavor Nambu - Jona-Lasinio model including
2SC, uniform chiral and non-uniform chiral phases was calculated within the mean-field approximation.

\section{Mean field and Nambu-Gorkov formalism}

We consider the NJL model at finite temperature:
\begin{equation}
\label{NJL}
H=\int_x\left\{\bar{\psi}(i\gamma^\nu\partial_\nu +\mu\gamma_0)\psi
+G\left[ (\bar{\psi}\psi )^2+(\bar{\psi}i\gamma_5\vec{\tau}\psi )^2\right]
+G^\prime (\bar{\psi}_ci\gamma_5\tau_2\lambda^A\psi ) (\bar{\psi}i\gamma_5\tau_2\lambda^A\psi_c )\right\}
\end{equation}
where $\psi $ is the quark field, $\psi_c=C\bar{\psi}^T$ the conjugate field
 and $\mu $ is the quark chemical potential.
The color, flavor and spinor indices are suppressed. The vector
$\vec{\tau}$ is the isospin vector of Pauli matrices and $\lambda^A$, $A=2,5,7$ are three color
antisymmetric $SU(3)$ group generators.
The integration $\int_x=\int_0^\beta d\tau\int d^3x $, where $\beta $ is the inverse temperature and
derivative operator $\partial_\nu = (i\partial_\tau ,\vec{\nabla})$. The coupling constant
$G$ describes the interaction in the isospin singlet, Lorentz scalar and the isospin triplet,
Lorentz pseudoscalar, quark-antiquark channels whereas $G^\prime$ describes the interaction in
the color $\bar{\mbox{\bf{3}}}$, flavor singlet, Lorentz scalar diquark channel. Both couplings
are treated as independent. In the momentum space a three momentum cut-off $\Lambda $
is introduced which finally defines the model.

Using the Hubbard-Stratonovich transformation one can perform partial bosonization of the
model which leads to the standard expression
\begin{eqnarray}
\label{NJL_bos}
H=\frac{1}{2}\int_x\left\{\bar{\psi}(i\gamma^\nu\partial_\nu +\mu\gamma_0+\sigma+i\gamma_5\vec{\pi}\cdot\vec{\tau})\psi
+\bar{\psi}_c(i\gamma^\mu\partial_\mu -\mu\gamma_0+\sigma+i\gamma_5\vec{\pi}\cdot\vec{\tau}^T)\psi_c \right. \\\nonumber
\left. +G^\prime (\Delta_A\bar{\psi}_ci\gamma_5\tau_2\lambda^A\psi  + \Delta_{A}^\ast\bar{\psi}i\gamma_5\tau_2\lambda^A\psi_c )
-\frac{\sigma^2+\vec{\pi}^2}{2G}-\frac{\Delta_A\Delta_{A}^\ast}{2G^\prime}\right\} .
\end{eqnarray}
We analyze the model in the mean field approximation within the following ansatz
\begin{eqnarray}
\label{ansatz}
&&\sigma = -M\cos\vec{q}\cdot\vec{x} \\\nonumber
&&\pi_a = -M\delta_{a3}\sin\vec{q}\cdot\vec{x} \\\nonumber
&&\Delta_A = \Delta\delta_{A2}
\end{eqnarray}
which takes into account three phases
\begin{itemize}
\item the homogenous chiral phase Ch when the wave vector $\vec{q}=0$, the gap parameter $\Delta =0$
and the constituent quark mass $M\neq 0$.
\item The non-uniform chiral phase NCh when both the wave vector $\vec{q}\neq 0$ and constituent mass $M\neq 0$
whereas the gap parameter $\Delta =0$.
\item The superconducting 2SC phase when $\vec{q}=0$, $M = 0$ and $\Delta\neq 0$.
\end{itemize}
The chiral phases and 2SC phase could coexist with each other, as actually is the case for moderate
densities, which was showed at $T=0$ \cite{sad2}. One can also consider other condensates as
for example the mixed states of $\pi^+\, ,\pi^-$ with the zero net charge \cite{dautbron, sad1}.
All conclusions given below which are based on the energy considerations does not change
because the model treats all chiral fields at the same level.

The finite temperature analysis would be performed using Matsubara formalism in the Nambu-Gorkov representation.
The partition function has a form
\be
Z=\int D\bar{\psi}D\psi D\bar{\psi}_cD\psi_c D\sigma D\vec{\pi}D\Delta_AD\Delta_{A}^\ast\exp H .
\ee
The mean field ansatz (\ref{ansatz}) is space dependent thus before integration over the fermionic fields one makes
rotation $\psi^\prime = \sqrt{U}\psi$, where $U=\exp (i\gamma_5\tau_3\vec{q}\cdot\vec{x})$.
Then after introduction of the Nambu-Gorkov (NG) basis $\chi^T=(\psi^\prime ,\psi_{c}^\prime )$ one arrives at the
mean field action
\begin{eqnarray}
\label{NJL_MF}
&&H_{MF}=\frac{1}{2}\int_x\left( \bar{\chi}S^{-1}(x,y)\chi -\frac{M^2}{2G}-\frac{|\Delta |^2}{2G^\prime}\right) ,\\\nonumber
&&S^{-1}(x,y)=\left[\begin{array}{cc}
i\gamma^\nu(\partial_\nu - \frac{1}{2}i\gamma_5\tau_3q_\nu ) + \mu\gamma_0-M & i\gamma_5\tau_2\lambda_2\Delta \\
i\gamma_5\tau_2\lambda_2\Delta^\ast & i\gamma^\nu(\partial_\nu - \frac{1}{2}i\gamma_5\tau_3q_\nu) -\mu\gamma_0-M
\end{array}\right] ,
\end{eqnarray}
where $q^\nu =(0,\vec{q})$. The bilinear form of $H_{MF}$ can be integrated over the fermionic fields with
the mean field result
\be
\label{ZMF}
Z_{MF}=\exp \left[-\int_x \left(\frac{M^2}{4G}+\frac{|\Delta |^2}{4G^\prime}\right) + \frac{1}{2}\ln\det S^{-1}\right] .
\ee
Determinant of the inverse propagator is calculated in the momentum space
\be
S^{-1}(K)=\int_K S^{-1}(x,y)\exp (-i K(x-y)) ,
\ee
where
\be
\int_K=T\sum_{n=-\infty}^\infty\int\frac{d^3k}{(2\pi )^3},\;\;\; K^\nu = (-i\omega_n ,\vec{k}),\;\;\;\omega_n=(2n+1)\pi T.
\ee
The inverse propagator $S^{-1}(K)$ for any given $K$ is a $48\times 48$ matrix in the color, flavor, Dirac and NG space.
However, it decays into two block-diagonal sub-matrices: $16\times 16$ matrix $S_{0}^{-1} $and $32\times 32$
matrix $S_{\Delta}^{-1}$. This simplification follows  because the diquark interaction
does not mixed the blue and red/green quarks. The calculation of these two determinants is
straightforward although rather tedious and leads to the result
\beq
\label{det}
&&\det S_{\Delta}^{-1} = \left[\prod_{i,k=\pm } (\epsilon_{ik}^2-K_{0}^2)\right]^4,\\\nonumber
&&\epsilon_{+ ,\,\pm}=\sqrt{(\mu + E_\pm)^2+|\Delta |^2},\;\;\;\epsilon_{- ,\,\pm}=\sqrt{(\mu - E_\pm)^2+|\Delta |^2} \\\nonumber
&&E_\pm = \sqrt{\vec{k}^2+M^2+\frac{\vec{q}^{\,2}}{4}\pm \sqrt{(\vec{q}\cdot\vec{k})^2+M^2\vec{q}^{\,2}}}
\eeq
where $\det S_{0}^{-1} = \det [S_{\Delta =0}^{-1}]$. After substitution of (\ref{det}) into (\ref{ZMF})
one gets the grand potential $\Omega =-T\ln Z_{MF}$ in the form
\beq
\label{Omega}
&&\frac{\Omega}{V} =  V_0
-4T\int^\Lambda\frac{d^3k}{(2\pi)^3}\sum_{i,k=\pm}\left( \ln \frac{1}{2}(1+\exp\frac{-\epsilon_{ik}}{T})
+\frac{1}{2}\ln \frac{1}{2}(1+\exp\frac{-\epsilon^0_{ik}}{T})\right) ,\\\nonumber
&&V_0 =\frac{M^2}{4G} +\frac{|\Delta |^2}{4G^\prime}
-2\sum_{i,k=\pm}\int^\Lambda\frac{d^3k}{(2\pi)^3} (\epsilon_{ik}+E_i)
+2\sum_{i=\pm}\int_{E_i\leq\mu}\frac{d^3k}{(2\pi)^3} (E_i-\mu) ,\\\nonumber
&&\epsilon^0_{ik}=\epsilon_{ik}(\Delta =0).
\eeq
In the limit of zero temperature the second term of the potential $\Omega $ vanishes. It means that the term $V_0$ describes
the zero temperature contribution. This term can be rewritten in the form useful for further calculation
\beq
\label{V0}
V_0=\frac{M^2}{4G} +\frac{|\Delta |^2}{4G^\prime} +\frac{M^2F^{2}_\pi \vec{q}^{\,2}}{2M^{2}_0}
- 12\int^\Lambda\frac{d^3k}{(2\pi)^3} E_{0} \\\nonumber
-2\sum_{i=\pm}\left[\int^\Lambda\frac{d^3k}{(2\pi )^3}(\sum_{k=\pm}\epsilon_{ik}
- 2E_{i})-\int_{E_i\leq\mu}\frac{d^3k}{(2\pi)^3} (E_i-\mu)\right]
\eeq
where the expansion in the power of the wave vector $\vec{q}$ was used as described in \cite{sad2}
and $M_0=0.301$ GeV is a constituent quark mass at zero density.
The quantity $E_0=E_i(\vec{q}=0)$ and $F_\pi = 93\;\MeV $ is a pion decay constant.

\section{Results}

The phase diagram describes the global minima of the potential $\Omega$ as
a function of temperature $T$ and chemical potential $\mu$. Further constraints
can be also imposed on the system as color and charge neutrality.
We do not include them in this paper because its influence on the competition
between chiral and 2SC phases is not quantitatively substantial. This problem would be
discussed at the end of this section. The thermodynamic potential $\Omega = V_0+V_T$ is minimized
with respect to $M$, $\vec{q}$ and $\Delta $. The zero temperature contribution
$V_0$ is given by equation (\ref{V0}) and $V_T$ is the finite temperature part
of expression (\ref{Omega}) (second term of $\Omega$). The model parameters
can be fitted to the values of the chiral condensate and the pion mass which give
$G=5.01$ GeV$^{-2}$ and the cut-off $\Lambda = 0.65$ GeV \cite{klev}.
The coupling constant $G^\prime$ was treated as independent from $G$.
The minima were found using standard Nelder-Mead Simplex algorithm.
The calculation was performed starting from three different points related
to three different phases. The global minimum was that of the lowest value of
the thermodynamic potential.

\begin{figure}
\centerline{\epsfxsize=9 cm \epsfbox{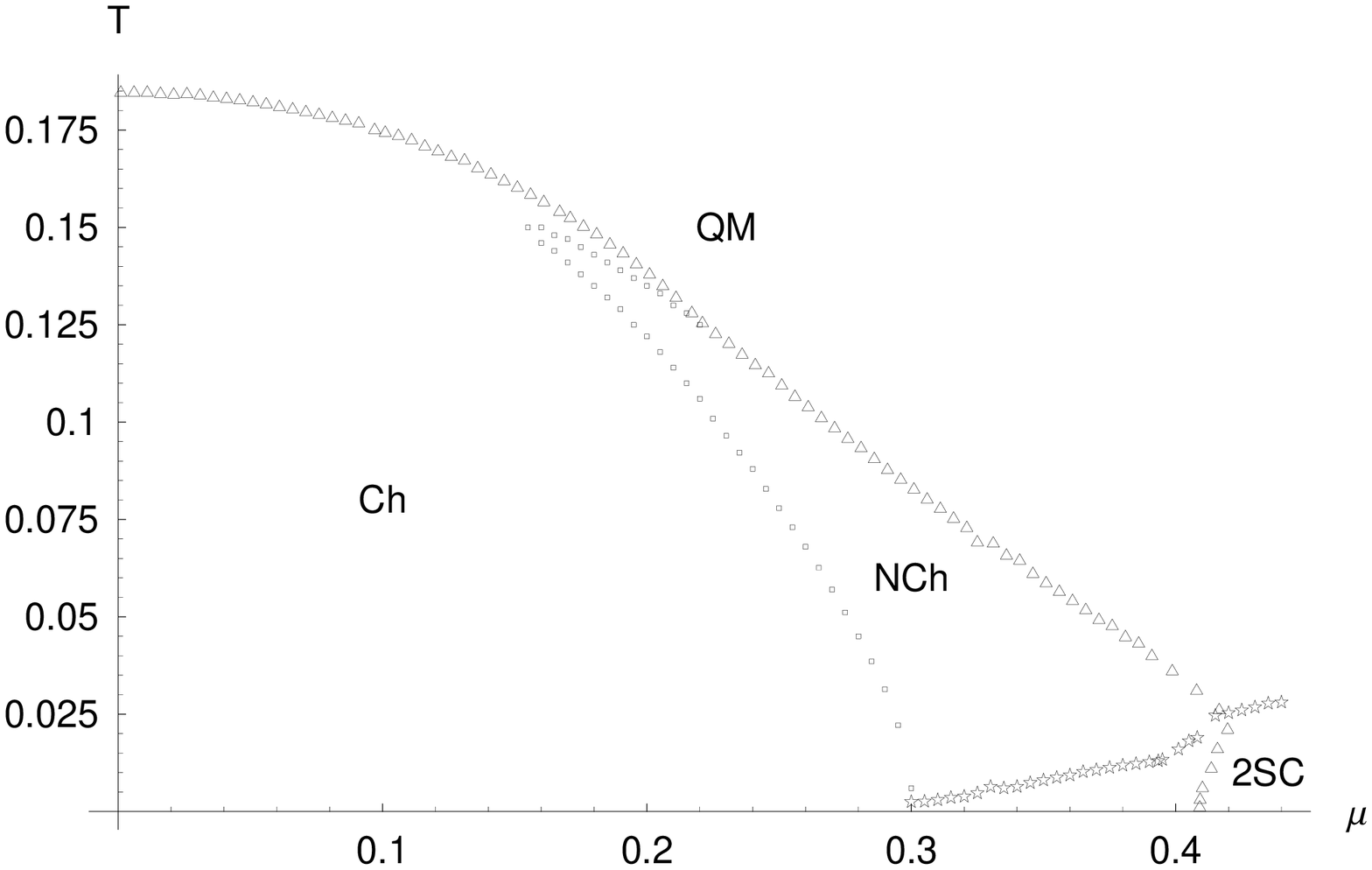} \epsfxsize=9 cm \epsfbox{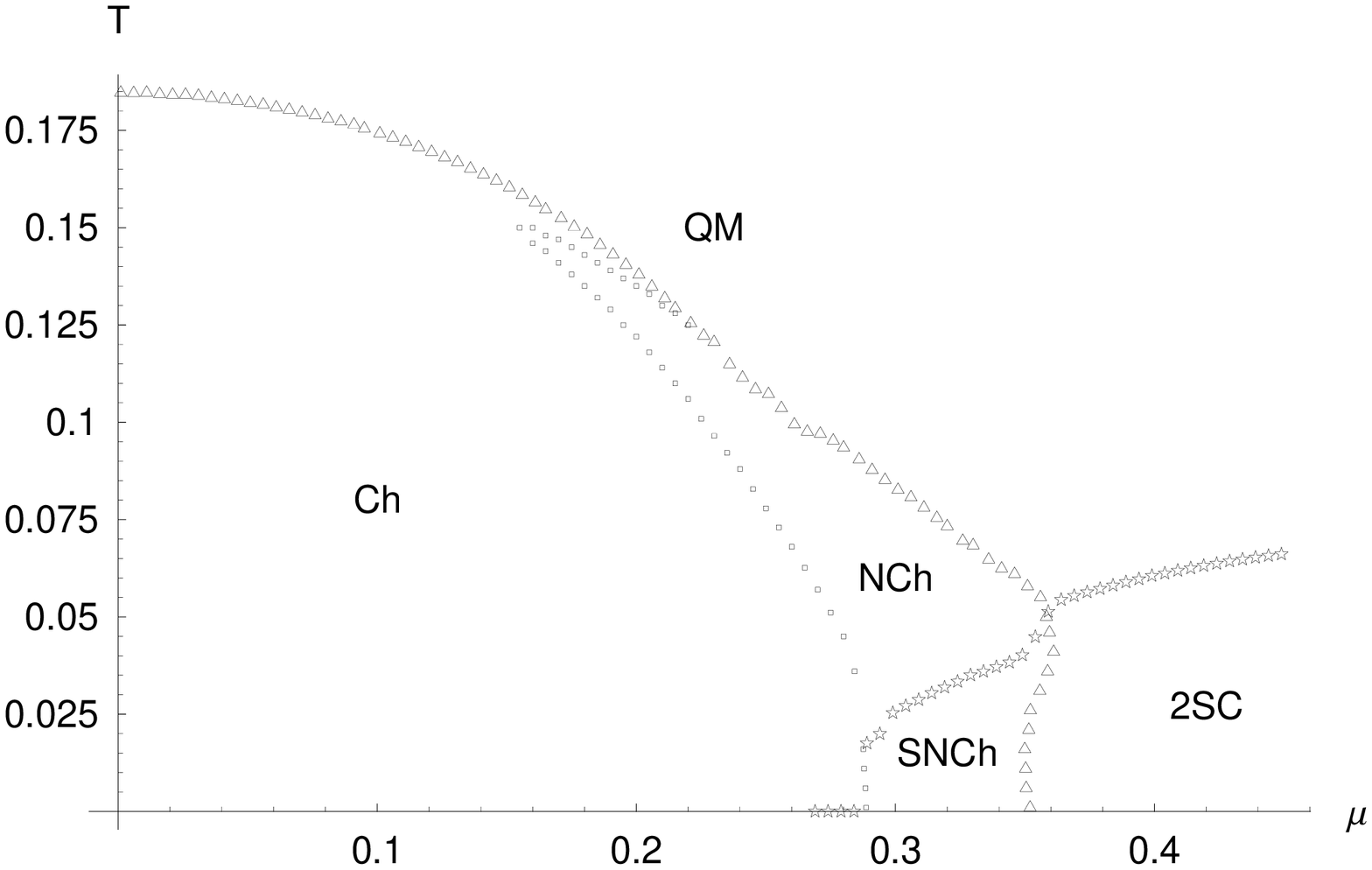}}
\caption{The phase diagram in the $T - \mu$ plain in GeV for the coupling constants
$G^\prime = G/2$ (left panel) and $G^\prime = 3G/4$ (right panel).}
\end{figure}

In the Fig. 1 the phase diagram for $G^\prime = G/2$ (left panel)
and $G^\prime =3G/4$ (right panel) are described. There are five different phases
\begin{itemize}
\item the quark matter phase (QM) with $M= 0,\,\vec{q}=0,\,\Delta=0$.
\item The chiral phase (Ch) $M\neq 0$ $\vec{q}=0,\,\Delta=0$.
\item The non-uniform chiral phase (NCh) $M\neq 0,\,\vec{q}\neq 0$, $\Delta=0$.
\item The mixed phase of color superconducting and non-uniform chiral phases (SNCh)
$M\neq 0,\,\vec{q}\neq 0,\,\Delta\neq 0$.
\item The color supercoducting phase (2SC) $M=0,\,\vec{q}=0$, $\Delta\neq 0$.
\end{itemize}

\begin{figure}[b]
\centerline{\epsfxsize=9 cm \epsfbox{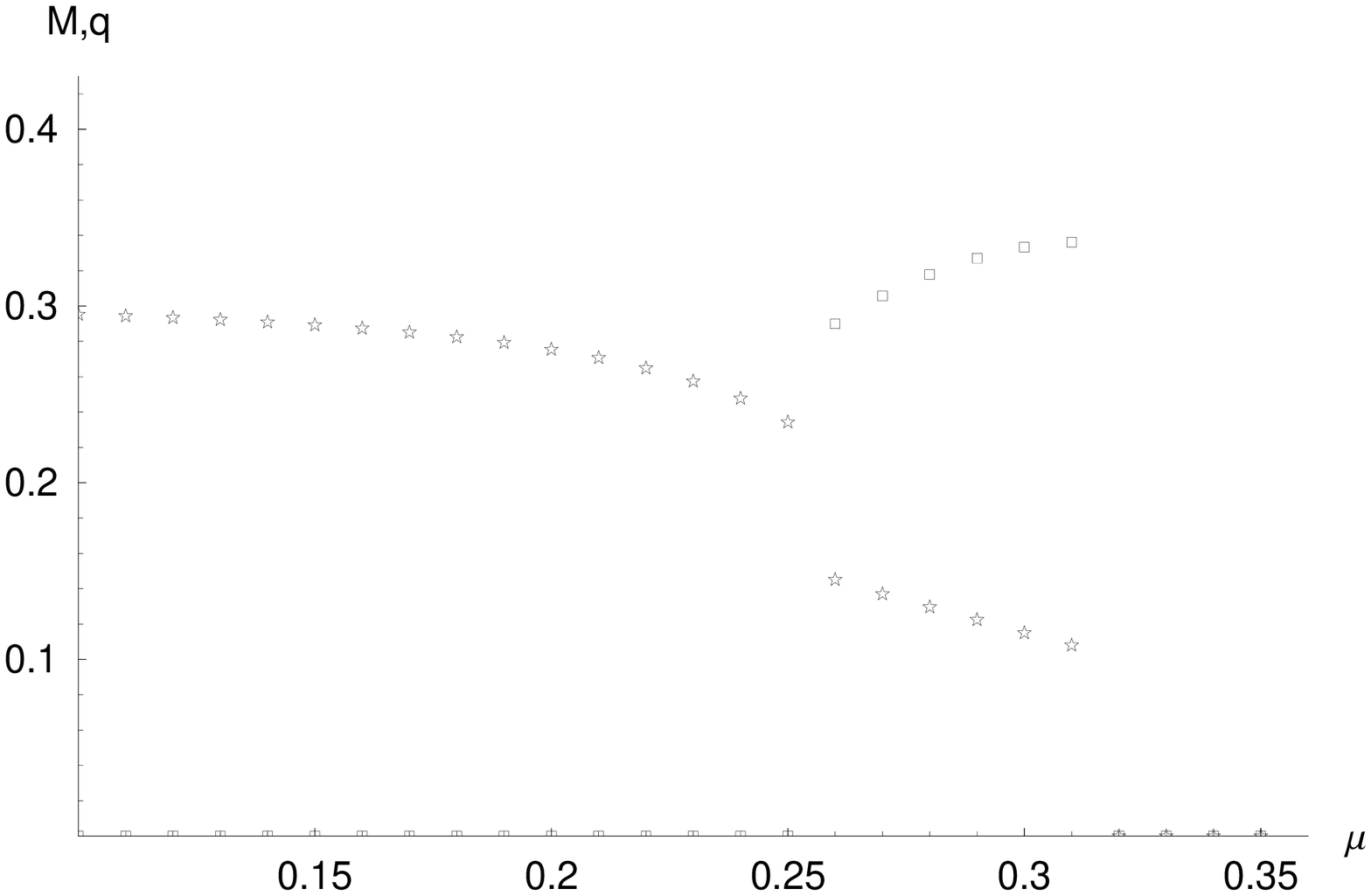} \epsfxsize=9 cm \epsfbox{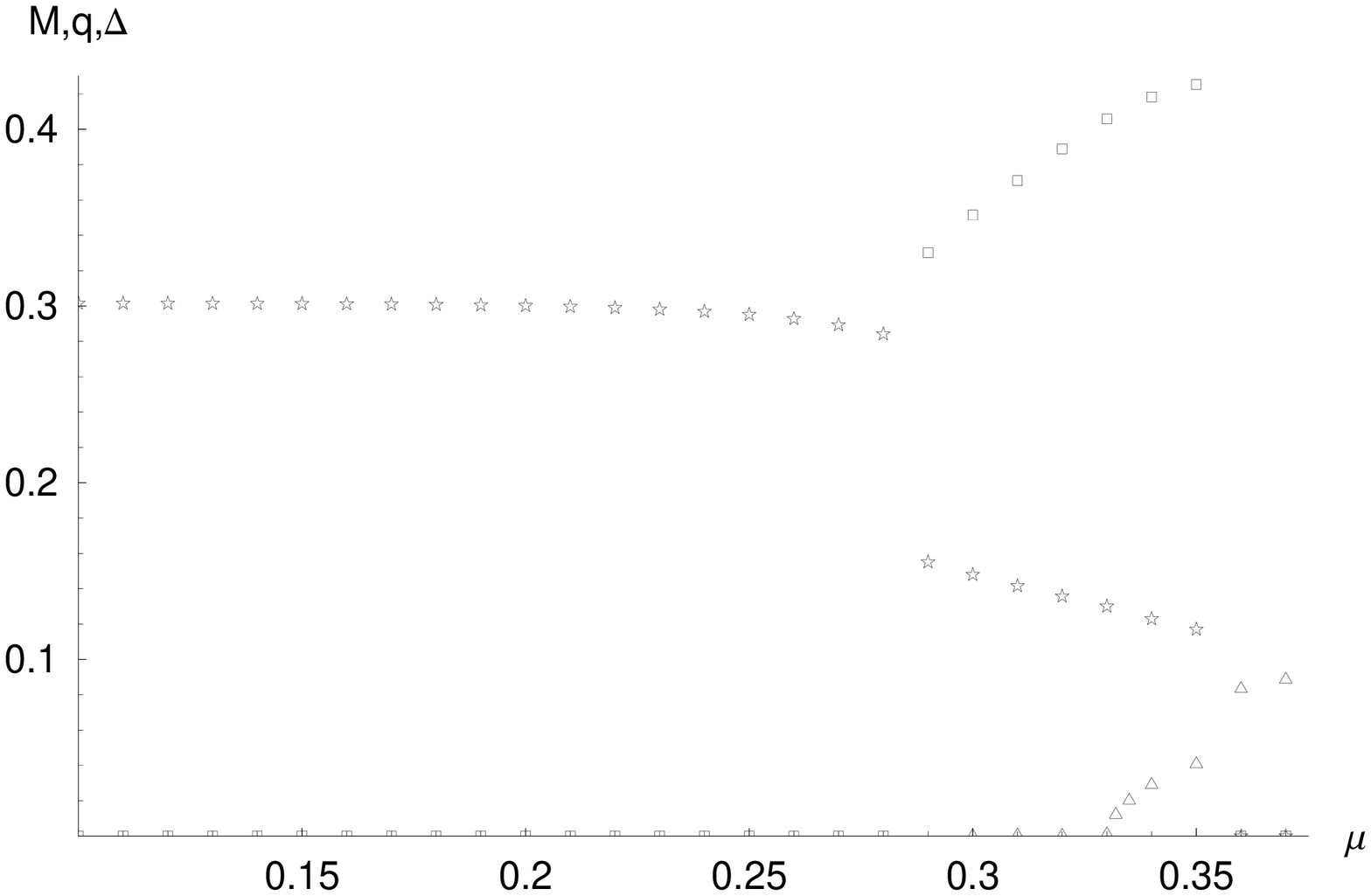}}
\caption{The value of the mass $M$ (stars), wave vector $q$ (boxes) and the gap parameter
$\Delta $ (triangles) for $T=0.075$ GeV (left panel) and $T=0.035$ GeV (right panel)
for coupling constants $G^\prime = 3G/4$.}
\end{figure}

Each of these phases differ by the order parameter and can be easily distinguished from
each other. The phase transition lines between Ch/NCh, QM/NCh, Ch/SNCh and 2SC/SNCh
are first order. This is connected to the fact that the entrance to the region of
non-uniform condensate is associated with the jump in the value
of the wave vector. This discontinues change entails the discontinues changes
in the mass and gap parameters (see Fig. 2). The QM/Ch, QM/2SC and NCh/SNCh lines are second
order phase transitions. This is nicely seen on the right panel of the Fig.2 which depicts
the values of order parameters at the temperature $T=0.035$ GeV as a function of the
chemical potential. First there is a 1st order phase transition from Ch into the NCh phase
around $\mu = 0.29$ GeV which is followed by the continuous phase transition into SNCh
phase around $\mu = 0.33$ GeV (rise of the gap parameter $\Delta$).
Eventually there is the 1st order transition into the 2SC phase at $\mu = 0.35$ GeV, where both mass and the wave vector
parameters vanish. The change of the character of the QM/Ch transition line
from the second to the first order takes place in the merge point around $(\mu ,\,T)=(0.221,0.125)$ GeV
where QM, Ch and NCh phases meets with each other.

The details of the phase diagram depends on the strength of the diquark
coupling constant. The lower value of $G^\prime$ the smaller part of the diagram is covered
with the pure 2SC phase. In the Fig. 1 it is seen that at low
temperature the SNCh/2SC phase boundary starts at $\mu\approx 0.409$ MeV for $G^\prime = G/2$
in compare to $\mu\approx 0.35$ MeV for $G^\prime = 3G/4$. However when
temperature grows both phase diagrams start to
resemble each other (also quantitatively). Such a behavior is expected because the
color superconductivity is rather low temperature phenomenon and does not
influence the phase structure at higher temperature. In particular
the region of NCh field is almost independent of the value of
$G^\prime$ coupling constant. The only difference is that SNCh phase
extends to the wider region. Also the melting temperature of the 2SC
phase is larger at the right panel compare to the left one.
This follows from the fact that the gap parameter takes
larger values when $G^\prime = 3G/4$.

The phase diagram at higher temperature ($T\approx 120$ MeV) has to be
treated with caution because of the corrections
from the pion loops. In particular the characteristic insert of the NCh phase
(the horn going to the left from the merge point of NCh/QM lines) possesses
the wave vector of a small value $|\vec{q}|\ll \pi T$ thus the temperature fluctuations
rather destroyed that part of the phase. Moreover the details of the phase diagram at higher
temperature certainly depends on the regularization procedure, as one can conclude comparing
Fig. 1 with similar from the paper \cite{sad1}. The difference between these two
approaches comes from the different way of the regularization of
the $V_T$ part of the thermodynamic potential. Nevertheless at the lower
temperature the results are robust within the expected accuracy
of the NJL model.

Finally one has to remember that 2SC phase describes the  colorful
strongly interacting matter. The color charge neutrality condition can be easily
accommodated through the additional color chemical potential \cite{neutral}.
The 2SC phase is suppressed under this condition thus as a result the (S)NCh phases extend to
the larger region of the diagram. However the change in the phase diagram structure
is not substantial. At zero temperature it differs only in details at the few per
cent level at most \cite{domino}.

The influence of the beta equilibrium condition on the
superconducting phases can result in the change of the phase structure of the color superconductor \cite{LOFF}.
Then one needs rather to compare the non-uniform chiral phase with the LOFF superconducting
phase. However, this task certainly requires different study and is outside the scope
of the present paper.

\section{Conclusions and future prospects}


In this paper the non-uniform chiral condensate NCh was compared
to the uniform 2SC phase (\ref{ansatz}) at finite densities and
temperatures. It was shown that the NCh phase exists in the temperature range between 20 - 100 MeV
above chemical potential $\mu\sim 250$ MeV. It also coexists with 2SC phase at
lower temperatures. The range of the phase coexistence depends on the relative strength of
the chiral and diquark couplings. For $G^\prime = G/2$ the density window
open at $\mu\approx 0.3$ GeV and lasts for about 100 MeV
whereas for $G^\prime = 3G/4$ it shrinks to about 50 MeV. The detailed phase
diagram can be found in the Fig. 1. It is worth to remind that NCh phase can have an important influence
on the neutron star structure and also can be a source of the strong
magnetic field \cite{dautbron,takahashi}. Then, the mixed state SNCh can have
very unusual features as composed of NCh and 2SC phases. This interesting expectation remains
to be revealed in some future work.

The light quarks are assumed to be massless in the model. The introduction of the non-zero
mass leads to the tilt in the chiral wave
as was shown in the low dimensional model \cite{thies}. There is not known exact solution
to this problem in 3+1 dimension. However, the effect has to be proportional to the current
quark masses, which are the smallest energy scales in the system and should not have
an important influence on energetic considerations. Nevertheless, the detailed analysis
remains as an open question.

Another important problem is the comparison of the NCh phase and
the color superconducting LOFF phase. This last phase appears naturally
under the condition that matter is at beta equilibrium in the interior of the compact
stars  \cite{alfraja}. It is also consider as a remedy for the chromomagnetic instabilities
in two flavor color superconductor \cite{giannakis}. At the first step
one can try the simplified version of the problem when only two wave
vectors for chiral and superconducting phase are assumed. The more general
ansatz would be more ambitious, however, much harder to solve. Nevertheless
the full analysis of the ground state at high baryon densities would
require the study of the non-uniform phases of the chiral as well as
diquark condensates.

\vskip0.25in
\centerline{\bf Acknowledgments} I would like to thank Piotr Bialas and
Jan Kotanski for useful discussions. The research was supported by the
MEiN grant 1P03B 045 29 (2005-2008).

\end{document}